# Anodic TiO$_2$ nanotube layers: why does self-organized growth occur - A mini review

Xuemei Zhou[a], [‡] Nhat Truong Nguyen[a], [‡] Selda Özkan[a], [‡] Patrik Schmuki*[ab]

[a]*Department of Materials Science WW4-LKO, University of Erlangen-Nuremberg, Martensstrasse 7, 91058 Erlangen, Germany*

[b]*Department of Chemistry, Faculty of Science, King Abdulaziz University, Jeddah, P.O. Box 80203, Jeddah 21569, Saudi Arabia*

[‡]These authors contributed equally.

* Corresponding author: P. Schmuki – schmuki@ww.uni-erlangen.de

**Abstract**

The present review gives an overview of the highlights of more than 10 years of research on synthesis and applications of ordered oxide structures (nanotube layers, hexagonal pore arrangements) that are formed by self-organizing anodization of metals. In particular we address the questions after the critical factors that lead to the spectacular self-ordering during the growth of anodic oxides that finally yield morphologies such as highly ordered TiO$_2$ nanotube arrays and similar structures. Why are tubes and pores formed - what are the key parameters controlling these processes?

Over the past decade, the formation and application of anodic, self-organized $TiO_2$ nanotube arrays grown from a Ti metal (as illustrated in Fig.1) and similar oxide structures have attracted wide scientific and technological interest [1-7]. Self-organization during anodization has been observed already more than 70 years ago for aluminum that, when anodized in acidic electrolytes, forms hexagonally ordered porous structures [8] – such ordering can peak in a virtually perfect arrangement, as demonstrated by Masuda et al. in 1995 [9], by a meticulous optimization of the electrochemical growth conditions. These alumina structures since then have been widely used for templating (i.e., to fill the pores by a secondary material to form nanorods or nanowires, in combination with stripping the template or not), as well as for numerous other applications - excellent overviews on growth and applications of porous alumina are available; see e.g. refs. [2-4,10,11].

In the past few years, however, research activities on ordered $TiO_2$ nanotube layers have in numbers of paper-output surpassed porous alumina; in the past decade, more than three thousand papers have been published dedicated to anodic $TiO_2$ nanotubes [1,5-7]. The main reason for this enormous interest is the anticipated impact of such nanotube layers in functional applications of titanium dioxide; such as dye sensitized solar cells [12-15], photocatalysis [16,17] (including water splitting for the generation of hydrogen, pollution degradation, or the reduction of $CO_2$), biomedicine [18-20] (biomedical coatings of implants, drug delivery systems), ion-insertion batteries, electrochromics, etc. [21-24] These applications are, to a large extent, based on a number of almost unique features of $TiO_2$ [1,25-28]: it is a semiconductor of a band-gap of 3.0 eV (rutile) - 3.2 eV (anatase), with a considerably large electron diffusion length (mainly anatase), relative band-edge positions suitable to trigger a wide range of photocatalytic reactions; the material is highly biocompatible, and shows considerably good ion intercalation properties. Many of these features can be exploited in a nanotubular form even more beneficially than in powder assemblies (e.g. directional charge transport, orthogonal carrier separation, optimized and directional diffusion profiles, defined membranes, biological cell interactions, possible quantum size and tubular core shell devices). A particular advantage of anodic nanotube arrays is that tubes are grown from their metallic substrate and thus can be used directly as back-contacted oxide electrode.

While a number of excellent literature reviews exist on the immense efforts that are directed towards an ever increasing control over growth conditions as well as exploiting the tube layers in applications [1,5-7], only relatively little work deals with the question after the origin of self-organization in anodic oxide formation processes [29-35]. It is thus the goal of the present review to particularly address this point and give an overview of factors dictating self-organization and tube formation.

First reports on self-ordered $TiO_2$ nanotube formation date back to 1984 and work by Assefpour-Dezfuly [36] and later by Zwilling in 1999 [37] who described that the use of dilute fluoride electrolytes in anodization of titanium can to lead to self-organizing oxide growth. Nevertheless, this first generation of nanotubes (including later works by Grimes [38] or Beranek [39]) used acidic, aqueous electrolytes, and led to tubes that showed considerable inhomogeneity in ordering, rough walls, and a tube length limited to < 1 μm.

Experimentally significant improvements were made by introducing pH mediation [40] as well as non-aqueous fluoride electrolytes [41-43] (mostly glycerol, ethylene glycol, DMSO, concentrated acids or ionic liquids). Together with voltage control and -alteration procedures, this allowed to establish smooth walled tube layers of several 100 μm thickness, bamboo or stack morphologies, branching, diameter control in the range from 10-800 nm, and the creation of single/double walled morphologies, for an overview, see e.g. refs [1,44,45]. It should be noted though that the as-formed tubes are amorphous and contain significant amount of fluorides and typically carbon, and for many applications need to be thermally crystallized to anatase or mixed anatase/rutile structures [46].

Furthermore, it is worth mentioning that using dilute fluoride electrolytes, similar aligned oxide tubes or pore-morphologies can be grown on a full range of other metals (Zr [47-49], Hf [50], Ta [51-53], Fe [54-56], Nb [57,58], Co [59], V [60]) and many alloys, see e.g. ref. [1] for an overview.

Factors dictating self-organization:

Classic findings on self-organization during anodic oxide growth mainly come from observations on ordered porous alumina, and were later confirmed or adapted to self-ordered $TiO_2$ nanotube growth:

i) Anodic oxide formation is a high field ion formation/transport process as illustrated in Fig.1c, where ion migration is controlled by the field across the oxide layer according to eq. $I = A \exp(BF) = A \exp(B \Delta U/d)$, where $I$ is the current, $\Delta U$ is the voltage across the oxide layer thickness d defining the electric field (F = ΔU/d) and $A$ and $B$ are experimental constants [61]. For a given voltage and an insoluble oxide the process is self-limiting (the oxide grows thicker, the field drops and eventually becomes too weak to aid further ion migration). The result is thus a compact layer of the field thickness d=f·ΔU, where f is the oxide growth factor [f($TiO_2$)= 2.5 nm/V], Fig. 1b, *i*). A prerequisite for an extended growth of anodic oxide is to establish an oxide formation/dissolution equilibrium. In other words, high field conditions and thus a constant ion-flux to form permanently new oxide can only be maintained if the steady-state oxide is thin enough. Thus a certain degree of "solubility" of the oxide in the electrolyte is needed to provide continuous oxide growth for building up 3D structures [1,31]. In the case of $TiO_2$ nanotubes, this is achieved by solvatizing $TiO_2$ as fluoride complexes, either by chemical dissolution of the oxide or by capturing arriving $Ti^{4+}$ species (ejected to the electrolyte) by complexation as $[TiF_6]^{2-}$(Fig. 1d). However, if the dissolution rate is too high no steady-state oxide layer is formed, but complete oxide dissolution occurs (electropolishing of Ti) (Fig. 1b, *ii*).

ii) Stress: In order to explain the initial formation of a distinct tube morphology (initiated by a curved or scalloped metal/oxide interface), mainly the buildup of compressive stress during the early phase of anodic oxide is considered to be a key element. Stress is created by plain volume expansion when converting metal to oxide (Pilling-Bedworth ratio) and additionally may be enhanced by voltage induced electrostrictive forces [37,62]. Nevertheless, if there was not a continuous dissolution (see Fig. 1c), this stress (or the induced rounding at the metal/oxide interface) would just be frozen-in in a compact oxide layer. Experimental findings [63] such as in Fig. 2a clearly confirm the build up of net compressive stress in the initial phase of $TiO_2$ anodization.

iii) Field effects: Once interface curvature and oxide dissolution have started, distinct spots will show enhanced localized dissolution due to field concentration effects (Fig. 2b) [64]. This is very well in line with work of Ono et al. [30] describing that a higher degree of order is obtained for more rapid growth (high current densities). In particular, current conditions close to breakdown allow ideal hexagonal ordering, as cells provide a maximum "outward expansion pressure" (Fig. 2c) for fast growth.

iv) Typically a certain set of experimental conditions such as $F^-$ content, $\Delta U$, pH, conductivity, $H_2O$ content leads to a maximum organization (as indicated in Fig. 2d, where the fastest growth results in highest degree of ordering [1,65] – note in Fig. 2d: if more drastic conditions (higher fluoride concentration or a higher voltage than in the optimized case) are used, the tube morphology is lost and turns either into electropolishing or sponge formation (Fig. 2e) [66]. Maximum ordering conditions may change during anodization, e.g. due to local acidification and ion accumulation at the tube bottoms [40,67]. Typically the influence of different electrochemical parameters correlates well with their influence on the steady-state oxide formation/dissolution equilibrium. Another factor that affects morphology is the current efficiency, i.e. how much of the current goes into side reactions – namely oxygen evolution. If oxygen evolution becomes too high, the tube morphology turns into a sponge morphology [40,67] (Fig. 2e). A morphology change can also be caused if too high voltages are applied – this then triggers (dielectric) breakdown events in the oxide and, in the mildest form, leads to holes in tube walls. Therefore the formation of a defined tube morphology is typically limited to a certain parameter range (as illustrated in Fig. 2e).

v) Once tubes are growing their diameters can be estimated, if one assumes that from point of oxidation, a hemispherically shaped oxide layer will grow with a radius of $r = f \cdot \Delta U$ (f is the anodic growth factor and $\Delta U$ is the applied potential). This simple approach is able to explain, to a large extent, the experimentally observed linear dependencies of pore/tube diameters on anodization voltage. (Of course, if non-aqueous, low conductivity electrolytes are used, the IR-drop in the electrolyte needs to be considered to determine the effective voltage [1]).

vi) With predefined pore initiation sites (such as induced pretexturing), the surface can guide tube growth (most typical techniques are multiple anodization, mould indenting, or FIB surface structuring) [68-71].

More recently, a number of fundamental quantitative approaches - namely using perturbation analysis [32-34] or numerical treatments [72] - deal with initiation and growth of self-ordered porous alumina and $TiO_2$ nanotubes. In the initial phase the main question is how a morphological instability such as a corrugated surface can be stabilized. A main conclusion from theoretical approaches (similar as in above empirical models) is that for a certain degree of interface stress and a certain flux of ions across the oxide interfaces (formation-dissolution equilibrium), instabilities of certain wavelength can be stabilized (an example is shown in Fig. 3a). Under certain assumptions these models are further able to quantitatively predict a dissolution range (expressed as oxide formation efficiency $\varepsilon_0$, as in Fig. 3b) that needs to be established to allow for self-organizing oxide morphologies. In Fig. 3b this is shown to hold for a range of experimental data of self-organized porous alumina and $TiO_2$ nanotube structures.

Nevertheless, one should point out that commonly the theoretically predicted correlation lengths often deviate from observed tube diameters [32,33], i.e. the currently used models may be applicable to predict the correlation length during initial layer formation (Fig. 3c) and may explain often observed "brainy" structures at tube bottoms at high anodization voltage (Fig. 3d). The diameter of tubes during actual growth may be better explained by a race for current, i.e. the elimination of smaller tubes (non fitting tubes) on the account of larger ones (Fig. 3e), until a uniform hexagonal filling of space is achieved. The diameter of the tubes then correlates well with the approach taken in Fig. 2f.

Another interesting phenomenon is that under self-organizing growth conditions, for porous alumina and $TiO_2$ nanotubes an unexpected lengthening during growth is observed [34,73,74]. This has been ascribed (based on convincing experimental observations) to stress-induced viscoelastic flow of anodic oxides [73,75] that may become possible under certain electrochemical conditions. This effect would then contribute to pushing oxide upwards the side walls of tubes/pores (as schematically illustrated in Fig. 1e).

The difference between a tube and a pore morphology

All above models hold essentially for porous alumina and $TiO_2$ nanotubes. In general, alumina anodized in acidic solution (or fluoride containing neutral solution) typically shows a nanoporous structure, while for $TiO_2$ virtually under all self-organizing conditions a tubular shape is formed [1,66]. It is important to realize that the only significant difference of the two morphologies is, if the cell boundaries of a porous structure can be etched under the applied electrochemical conditions. As illustrated in Fig. 1d in the case of $TiO_2$ nanotube formation, the small fluoride ions can easily compete with $O^{2-}$ inward migration during oxide growth [76]. As a result, fluoride ions accumulate at the metal/oxide interface and move upward with tube growth as an outer cladding (decorating the hexagonal cell walls) [1,77]. Ti-fluoride compounds are easily dissolved in $H_2O$ (Fig. 1f) - thus the fluoride rich locations between $TiO_2$ tubes (or cells) are dissolved. Due to the comparably longer etching of the tops, the resulting tubular morphology is more apparent at the top of the structure (Fig. 1a), while the bottom often still shows a tight hexagonal and porous arrangement (Fig. 1a, inset). In case of porous alumina, even formed in fluoride electrolytes [78], a porous morphology is obtained, as Al-fluorides are only slightly soluble in $H_2O$.

For $TiO_2$ the tubular shape is observed under a wide range of conditions (Fig. 2e). Only if the water content is drastically reduced, the cell boundaries are not easily etched and porous layers can be observed [79]. It should also be pointed out that the ratio of growth speed of the tubes versus the tube splitting speed (that is how fast the chemical etching of the intercell fluorides follows the growing tubes, indicated in Fig. 1f) strongly affects the tube morphology [1]. If growth is faster than splitting, smooth tube walls are obtained, in the opposite case corrugated tube walls (bamboo type walls) are obtained.

Specific other morphologies and factors

Above growth models and findings hold for close packed hexagonal arrangements, that is, if the density of pore/tube initiation sites that is naturally provided or intentionally created under certain experimental conditions is sufficiently high. If initiation occurs with a larger spacing, one may observe individual tubes that can be considerably separated from each other [43]. Typically these tubes can grow significantly in diameter and are embedded in a matrix of mesoporous oxide. This oxide is easier soluble in the electrolyte and thus leads finally to an apparent larger spacing between tubes (Fig. 3f). In the light of above arguments, the formation of such morphologies may be aided by a fast tube splitting rate with conditions that allow for sponge formation in between tubes ($O_2$ side reaction).

Another aspect, frequently overlooked in literature, is that anodic oxide layers often consist of a double layer structure (oxide that is formed by $O^{2-}$ inward migration and oxide formed by $Ti^{4+}$ outward migration). This leads to a twin-wall morphology of $TiO_2$ nanotubes [46]. Typically the inner shell of $TiO_2$ nanotubes is of inferior quality. Only under specific conditions [44,45] or using etching treatments this inner shell can be removed, and so-called "single-walled" tubes can be generated.

Concluding remarks

While in the present overview many aspects of the fascinating growth and structural features of $TiO_2$ nanotubes and similar oxide structures could only be touched on the surface and in a referencing manner, we hope to provide interested readers i) with a guideline through the wide literature to the topic, and ii) more specifically, to give a comprehensive overview of critical factors affecting self-organizing anodization.

Acknowledgements

The authors would like to acknowledge ERC, DFG and the Erlangen DFG cluster of excellence for financial support.

**Figure Captions**

Figure 1. a) SEM and TEM images of a typical TiO$_2$ nanotube structure (upper inset shows bottom view, lower inset TEM cross-section). Reproduced with permission from [1]. Copyright 2011 Wiley VCH Verlag GmbH & Co. b) Electrochemical anodization resulting in different oxide morphologies: for optimized solubility of oxide in electrolyte self-organized pores/tubes can be observed. c) Illustration of high field oxide formation process. At metal/oxide interface Ti$^{4+}$ is formed and migrates outward; O$^{2-}$ extracted from water at the oxide/surroundings interface migrates inward. Oxide can be formed e.g. at metal/oxide interface by reaction of arriving O$^{2-}$ with Ti$^{4+}$. Without oxide dissolution the film is compact and of a final thickness. d) High field oxide formation in the presence of fluoride ions, a steady-state is established between oxide formation at inner interface and dissolution at outer interface (due to dissolution/complexation of Ti$^{4+}$ as TiF$_6^{2-}$). Additionally rapid fluoride migration leads to accumulation of fluoride at the Ti/TiO$_2$ interface. e) Initial tube growth under the mutual effect of steady-state oxide formation/dissolution and compressive stress shaping oxide and fluoride-rich layer into roundings. f) Formation of a tube morphology by chemical dissolution by H$_2$O of fluoride-rich layer between hexagonal cells.

Figure 2. a) Stress built up at oxide/metal interface during anodization. Example shows stress·thickness vs. thickness curves obtained upon anodization of Ti in HNO$_3$ at various current densities. Reproduced with permission from [63]. Copyright 2006 The Electrochemical Society. b) Field concentration effects in anodic nanotube bottoms. c) Schematic representation of enhanced hexagonal self-organization at high current density, i.e. higher electric field. Reproduced with permission from [30]. Copyright 2005 Elsevier B.V. d) TiO$_2$ nanotube layer thickness as a function of the applied potential and the fluoride concentration with an optimized self-organization at highest growth rates. Reproduced with permission from [65]. Copyright 2007 Wiley-VCH Verlag GmbH & Co. e) Existence diagram for various TiO$_2$ morphologies (nanotubes, pores, sponge structures as well as the breakdown and electropolishing region) dependent on anodization parameters. Reproduced with permission from [66]. Copyright 2010 Wiley-VCH Verlag GmbH & Co. f) Oxide formation at a nucleation spot resulting in hemispherical oxide dome formation (illustrating voltage dependence of TiO$_2$ nanotube diameters).

Figure 3. Results from linear stability analysis according to [31]. a) Effect of different oxide formation efficiencies $\varepsilon_0$ on dispersion curves for growth of anodic Al$_2$O$_3$ at electric field of 0.8 V/nm. Anodic film is stable at $\varepsilon_0 > 0.7$ and shows an unstable region at $\varepsilon_0 < 0.66$; in between the perturbation wavelength decides on stability and formation of nanotubes/pores in certain regime. b) Comparison of measured oxide formation efficiencies $\varepsilon_0$ with predicted limits. Reproduced with permission from [31]. Copyright 2012 Macmillan Publishers Ltd. c) TiO$_2$ nanotubes with initiation layer on the top (note that the bottom of the initiation layer shows regular short wavelength corrugation). d) "brain" morphology

on bottom of anodic $TiO_2$ nanotubes grown at high currents (bottom shows short wavelength corrugation). e,f) Schematic illustration for the growth of $TiO_2$ nanotubes with e) high and f) low initiation nucleation density leading to growth of small tube exclusion and dense hexagonal packing (e) or single tube bulbs in a porous matrix (f), respectively.

**Figure 1.**

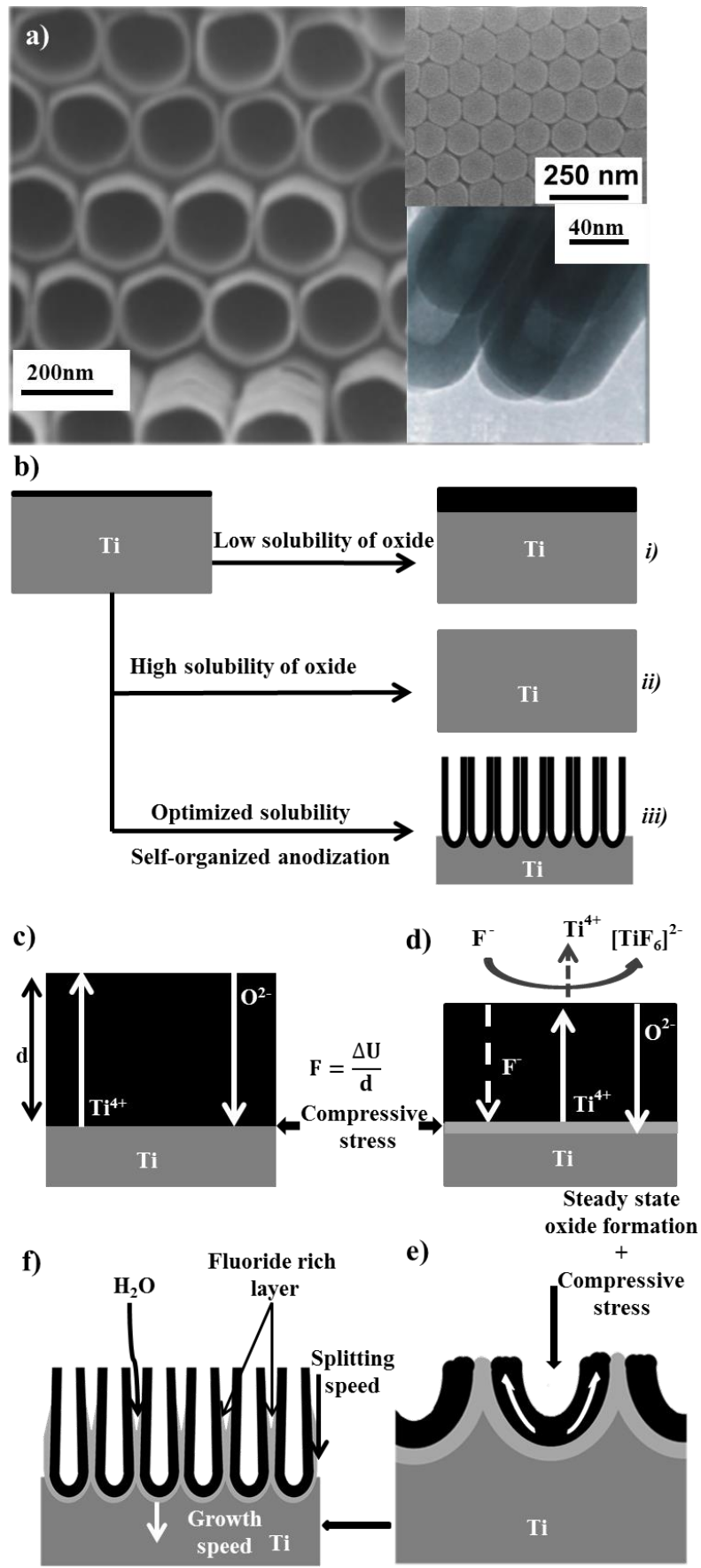

**Figure 2.**

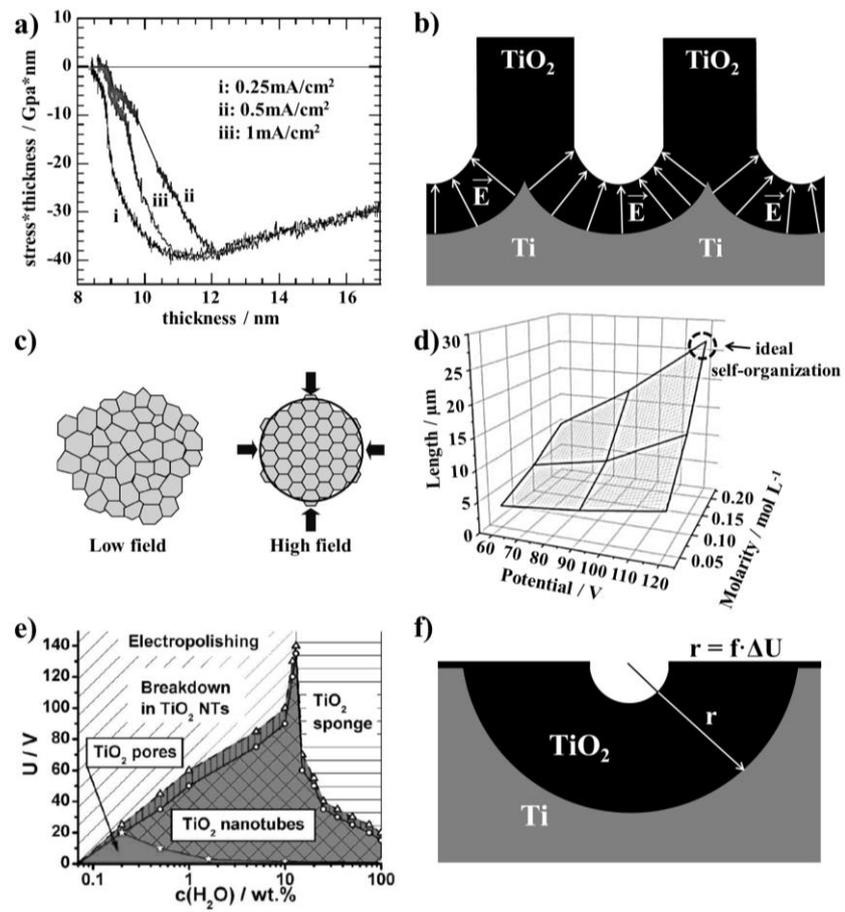

**Figure 3.**

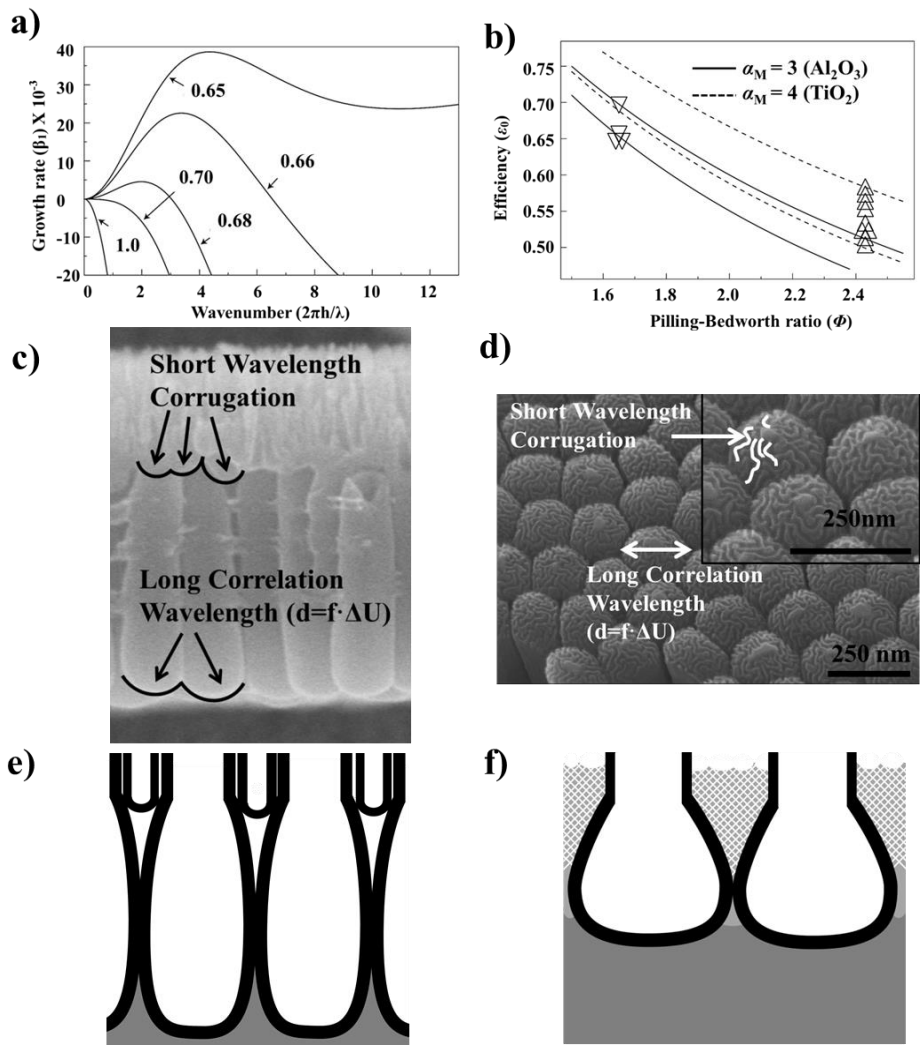